\documentclass[useAMS,usenatbib]{mn2e}
\usepackage{natbib}
\usepackage{graphicx}
\newcommand{\msun}{\mbox{$\,{\rm M}_\odot$}}

\title[The Curious Case of Palomar 13]{The Curious Case of Palomar 13: The Influence of the Orbital Phase on the Appearance of Galactic Satellites}

\author[A.H.W. K\"upper, S. Mieske and P. Kroupa]{Andreas
  H.W. K\"upper$^{1,2}$\thanks{E-mail: \mbox{akuepper@astro.uni-bonn.de} (AHWK); \mbox{smieske@eso.org} (SM); \mbox{pavel@astro.uni-bonn.de (PK)}}, Steffen Mieske$^{2}$ and Pavel Kroupa$^1$\\
$^{1}$Argelander Institut f\"ur Astronomie (AIfA), Auf dem H\"ugel 71, 53121 Bonn, Germany\\
$^{2}$European Southern Observatory, Alonso de Cordova 3107, Vitacura, Santiago, Chile}

\begin{document}

\date{Accepted \ldots. Received \ldots; in original form \ldots}

\pagerange{\pageref{firstpage}--\pageref{lastpage}} \pubyear{2010}

\maketitle

\label{firstpage}

\maketitle

\begin{abstract}
We investigate the dynamical status of the low-mass globular
cluster Palomar~13 by means of $N$-body computations to test whether
its unusually high mass-to-light ratio of about 40 and its peculiarly
shallow surface density profile can be caused by tidal
shocking. Alternatively,  we test -- by varying the assumed proper motion -- if the orbital phase of Palomar~13 within its orbit about the Milky Way can influence its appearance and
thus may be the origin of these peculiarities, as has been suggested
by \citet{Kuepper10b}. We find that, of these two scenarios, only the latter can explain the observed mass-to-light ratio and surface density profile. We note, however, that the particular orbit
that best reproduces those observed parameters has a proper motion
inconsistent with the available literature value. We discuss this
discrepancy and suggest that it may be caused by an underestimation of the
observational uncertainties in the proper motion determination. We demonstrate that Palomar ~13 is most likely near apogalacticon, which makes the cluster appear supervirial and blown-up due to orbital compression of its tidal debris. Since the satellites of the Milky Way are on average closer to apo- than perigalacticon, their
  internal dynamics may be influenced by the same effect, and we
  advocate that this needs to be taken into account when interpreting
their kinematical data. Moreover, we briefly discuss the influence of a possible binary population on such measurements.
\end{abstract}

\begin{keywords}
galaxies: kinematics and dynamics -- galaxies: star clusters -- globular clusters: individual: Palomar 13 -- methods: $N$-body simulations \end{keywords}

\section{Introduction}\label{Sec:Introduction}
There are many objects on the sky, especially in the halo of the Milky
Way (MW), whose nature is not clear to us. Some of those objects are hard
to address observationally, and for others there is just no conclusive
theoretical explanation. 

In fact, simply by looking at a
colour-selected sample of stars within a region of the sky it is
sometimes not easy to determine the true extent of a stellar system,
mostly since it lacks a clear cut-off in its surface density profile. The same holds true
for the determination of its velocity dispersion through a sub-sample
of stars with readily measured radial velocities. These
  uncertainties typically result in discussions and speculations
about a best-fitting density profile as well as a system's true tidal
radius (e.g. \citealt{King66, Elson87, McLaughlin05}), and also about
the true mass-to-light ratios of such systems (e.g. \citealt{Kroupa97,
  Mieske08}).

Some of these uncertainties arise from peculiar surface density profiles. That is, even though many objects in the Milky Way halo are well limited and show a well defined surface density profile with a slope of about
$R^{-4}$ in the region of the tidal radius, some objects obey shallow
surface density profiles in the outskirts, having slopes of about -1
to -2, like for example the MW globular clusters Palomar 5
\citep{Odenkirchen03}, NGC 5466, M 15, M 53, M 30, and NGC 5053
\citep{Chun10}, AM 4 \citep{Carraro07}, Whiting 1 \citep{Carraro09},
and NGC 1851 \citep{Olszewski09}. The latter furthermore seems to be
surrounded by a 500 pc halo of stars whose origin is unknown up to
now.

Other uncertainties arise from unusual mass-to-light (M/L)
  ratios of some stellar systems. While most globular clusters show mass-to-light ratios of
  1-2, Ultra-Compact Dwarf galaxies (UCDs) have higher M/L by a
  factor of about two, whereas dwarf spheroidal galaxies even show
  values of up to $10^3$ \citep{Dabringhausen08, Geha09}. These differences
  are usually ascribed to different dark matter contents,
catastrophic tidal heating by gravitational shocks, a variation of the
IMF, tidally reshaped stellar phase-space distribution functions, contaminations from stellar streams in the MW halo, or alternative gravitational theories (e.g. \citealt{Kroupa97}, \citealt{Gilmore07}, \citealt{Simon07}, \citealt{Mieske08}, \citealt{Angus08}, \citealt{Niederste09}).

The low-mass Galactic globular cluster Palomar 13 is a stellar system
which shows both, an unclear extent due to a shallow surface density
profile, and a high velocity dispersion resulting in a mass-to-light
ratio of about 40 \citep{Siegel01, Cote02}. Further details on this
cluster are presented in Sec.~\ref{Sec:Palomar 13}. In this
  investigation we demonstrate by means of $N$-body calculations how
  these observational results can be explained without the need for dark
matter, tidal heating, binaries or changes in the law of gravity. To this end we compute models of Palomar 13 on various orbits about
  the Galaxy that are consistent with its present-day distance and
  radial velocity with respect to the Sun. We show how different such
a stellar system can appear in different phases of its orbit. Details
on the models are described in Sec.~\ref{Sec:Models}. The results of
these computations and the mock observations in which we show
 how this cluster may appear when observed with an 8m-class
telescope, are shown in Sec.~\ref{Sec:Results}. Sec.~\ref{sec:discussion} is a short discussion on the plausibility of our findings. Finally, in Sec.~\ref{Sec:Conclusions} we give a short summary and conclusions.

\section{Palomar 13}\label{Sec:Palomar 13}
\begin{figure}
\includegraphics[width=84mm]{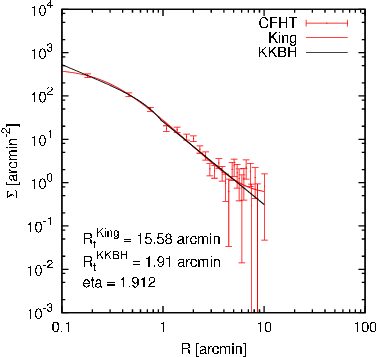}
  \caption{Surface density profile of Pal~13 as obtained with CFHT by
    \citet{Cote02}. Fitted to the profile are a King template
    \citep{King62} and a KKBH template \citep{Kuepper10b}. Given in
    the figure are the template values for the fitted tidal radius for
    both templates, $R_t^{King}$ and $R_t^{KKBH}$ respectively, and
    for the extra-tidal slope, $\eta$ (eta). The shallow slope at
    large radii of $\eta = 1.9$ influences the fit of the King
    template such that it yields a significantly larger value for the
    tidal radius as does the KKBH template. 1 arcmin corresponds to about 7 pc at the assumed distance of Pal~13.}
  \label{sdpCFHT}
\end{figure}
Palomar~13 is an old and metal poor Galactic globular cluster. From
isochrone fits to its colour-magnitude diagram, \citet{Cote02} find
Pal~13 to be about 13-14 Gyr old, and from spectroscopy they derive a
metallicity of $[\mbox{Fe/H}] = -1.9 \pm 0.2$. Moreover, it is among
the faintest objects listed in the Harris catalogue of Milky Way
globular clusters \citep{Harris96}. With an estimated mass of about
3000 $\msun$ (assuming a mass-to-light ratio of unity) it is one of
the least massive globular clusters of the Galaxy
\citep{Cote02}. Furthermore, \citet{Siegel01} argue with proper motion
measurements that Pal~13 is on an inclined, highly eccentric orbit
about the Milky Way. Thus, it most probably was subject to strong
tidal disruption during the last few Gyr. Indeed, observations show further peculiarities about this specific cluster. First, the cluster
shows an unusually high velocity dispersion and therefore a very high
mass-to-light ratio, and second, its surface density profile differs
from the ones of most other Milky Way globular clusters:
\begin{enumerate}
\item Corresponding to data by \citet{Cote02}, which we mainly use to
  compare with our computations, the cluster is located at Galactic
  longitude of $\ell = 87.^\circ 1$ and Galactic latitude of $b =
  -42.^\circ 7$. Its distance from the Sun is $R_{Sun} =
  24.3^{+1.2}_{-1.1}$ kpc, placing it at a distance of $R_{GC} =
  25.3^{+1.2}_{-1.1}$ kpc from the Galactic centre. Pal~13's radial
  velocity was determined by C\^ot\'e et al.~to be $V_r = (24.1 \pm
  0.5)$ km/s using spectroscopic data of the High Resolution Echelle Spectrometer at the Keck telescope. In the same investigation, its internal radial
  velocity dispersion was found to be $\sigma_r = (2.2 \pm 0.4)$ km/s
  from a sample of 21 stars located within the cluster's inner 2
  arcmin. Within their best estimate of Pal 13's tidal radius of 26
  arcmin, C\^ot\'e et al.~furthermore measured an absolute magnitude
  of $M_V = -3.8$ mag. Assuming Pal~13 being in virial equilibrium,
  this would imply a mass-to-light ratio of $M/L =
  40^{+24}_{-17}$. C\^ot\'e et al.~suggest that this unusually high
  velocity dispersion could be the consequence of either a
  catastrophic heating during a recent perigalacticon passage or the
  presence of a dark matter halo.\\
\item Fig.~\ref{sdpCFHT} shows the surface density profile as obtained
  with the 3.6 Canada-France-Hawaii telescope (CFHT) by
  \citet{Cote02}. The profile shows a shallow slope $\eta \simeq -2$ out to large radii (10
  arcmin correspond to about 70 pc at the distance of Pal~13), markedly different to the much
steeper slope $\eta \simeq -4$ found in most other surface density profiles of globular
  clusters (compare, e.g., \citealt{McLaughlin05}). This large extent
  of Pal~13 can be interpreted in two ways. First, the stellar
  population at large radii can be part of the cluster such that
  Pal~13 would be a very low concentrated cluster with a large tidal
  radius of about 26 arcmin [180 pc] \citep{Cote02}, or, second,
  Pal~13 can be interpreted as having a very pronounced tidal debris
  and the cluster itself having a significantly smaller tidal radius
  of about 3 arcmin [20 pc] \citep{Siegel01}. To check these
  oppositional proposals  for consistency with theoretical expectations,
  we can make a first estimate of Pal~13's true tidal radius, $R_t$,
  using
\begin{equation}
R_t = \left( \frac{GM}{2\Omega^2}\right)^{1/3}, 
\end{equation}
where $G$ is the gravitational constant, $M$ is Pal~13's present-day
mass, and $\Omega$ its angular velocity on its orbit about the Milky
Way \citep{Spitzer87}. This yields
\begin{equation}\label{eq:Rt}
R_t \simeq 43.7\,\mbox{pc}.
\end{equation}
when we assume that Pal~13 has a mass of about $3000\msun$ and is on a
circular orbit with an orbital velocity of about $V_{orb} = 220$ km/s
at $R_{GC} = 25.3$ kpc. In fact, Pal~13 is more likely to be on an
eccentric orbit and hence may rather have a present-day tidal radius
of about 50-100 pc when we take into account that its true angular
velocity is likely to be lower than that of a circular orbit. Thus, the theoretically expected range for Pal 13's tidal radius does not agree with either of the two observational estimates of C\^ot\'e et al. and Siegel et al.

\end{enumerate}

\subsection{Orbit}
\begin{figure}
\includegraphics[width=84mm]{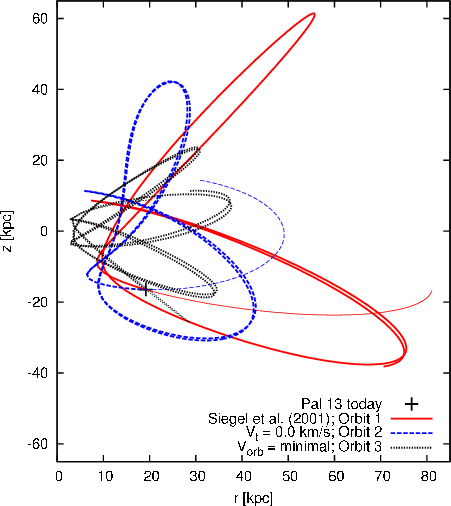}
  \caption{Orbit of Pal~13, as determined using the proper motion
    measured by \citet{Siegel01} and the radial velocity measured by
    \citet{Cote02}, referred to as \textit{orbit~1}, shown in the
    meriodal plane representation, where $r$ depicts the radial
    distance from the Galactic centre within the Galactic disk and $z$
    is the height above the Galactic disk. Shown are the last 3 Gyr
    and 0.5 Gyr into the future. Also shown is the resulting orbit
    after setting the transverse velocity, $V_t$, to zero, such that
    the cluster has only the (more precisely determined) radial
    velocity measured by C\^ot\'e et al.~(\textit{orbit~2}). The black
    dotted line depicts the orbit with the transverse velocity chosen
    such that it minimizes Pal 13's orbital velocity, $V_{orb}$,
    referred to as \textit{orbit~3}.}
  \label{orbits}
\end{figure}
The hypothesis that the above peculiarities are caused by tidal
effects is supported by measurements of \citet{Siegel01}, who, using
CCD photometry and 40 years older photographic plates, find Pal~13's
proper motion to be $\mu_\alpha\cos\delta = (2.30 \pm 0.26)$ mas/yr
and $\mu_\delta = (0.27 \pm 0.25)$ mas/yr. In Fig.~\ref{orbits} the
orbit of Pal 13 within the Milky Way is shown in red solid lines
  for the case of combining the proper motion of Siegel et al.~with
the radial velocity of \citet{Cote02}. Note that we will refer to this
orbit as \textit{orbit~1} throughout the text. The plotted line
  corresponds to the orbit integrated within an \citet{Allen91} Milky
  Way potential for the last 3 Gyr and 0.5 Gyr into the
  future. Distance and orbital motion of the Sun where taken from
\citet{Dehnen98}, that is, the Sun is located at a Galactocentric
distance of $x = 8$ kpc, the orbital velocity of the local standard of
rest (LSR) is 220 km/s in y-direction, and the Sun moves with respect
to the LSR with $V_x = 10.0$ km/s, $V_y = 5.3$ km/s and $V_z = 7.2$
km/s.

As we can see from Fig.~\ref{orbits}, corresponding to these measurements, Pal~13 has a very elliptical orbit with eccentricity
\begin{equation}
\epsilon = \frac{R_{apo}-R_{peri}}{R_{apo}+R_{peri}} = \frac{82.8-11.1}{82.8+11.1} = 0.76,
\end{equation} 
where $R_{apo}$ is its apogalactic distance and $R_{peri}$ its
perigalactic distance, respectively. Therefore, with a Galactocentric
distance of 25.3 kpc and the proper motion as measured by \citet{Siegel01}, Pal 13 is today quite close to its
perigalacticon. Because of this, \citet{Siegel01} and \citet{Cote02} suggest
that the high velocity dispersion of Pal 13 and the shallow slope of
its surface density profile at large radii may well be due to the last
pericentre passage which may have heated the cluster violently and may
have caused a rapid expansion and/or an overspilling over the tidal
boundaries.

A comprehensive $N$-body investigation
of low-mass globular clusters on eccentric orbits however shows that
pericentre passages at such great galactocentric distances barely
cause violent mass loss or rapid expansion \citep{Kuepper10a}. Furthermore,  a
follow-up investigation showed that the surface density profiles (a)
cannot be used to draw conclusions on the (theoretical) tidal radius
of a cluster, and (b) only show shallow slopes at large radii when the
cluster is close to reaching its apogalacticon \citep{Kuepper10b}. The
latter is due to the tidal tails of the cluster which get stretched
and compressed along the orbit. That is, if a cluster and its tails
move from apogalacticon to perigalacticon they get accelerated and
stretched, whereas from perigalacticon to apogalacticon they get
decelerated and compressed.

If the shallow slope in the surface density profile of Pal~13 is
indeed due to the effects described in \citet{Kuepper10a, Kuepper10b}, then Pal 13 would need to be currently in a
position close to apogalacticon. This, however, disagrees with the
position in its orbit (\textit{orbit~1}) derived from the proper
motion measured by Siegel et al. and that puts Pal 13 close to
perigalacticon (Fig. \ref{orbits}). In an attempt to resolve this
  contradiction, we will in the following define two further test
  orbits which differ in proper motion from the Siegel et
  al. estimate. We will then fully integrate the dynamical evolution
  of Pal 13 for all three orbits to compare the results with the
  observed surface brightness profile, luminosity, and radial velocity
  dispersion (Sec. \ref{Sec:Results}). The underlying assumption for this procedure
  is that the proper motion measurements are the most uncertain
  observed cluster properties.

We first force the cluster's transverse velocity
  (proper motion) to zero, while keeping the radial velocity. In Fig.~\ref{orbits} we can see
that this results in an orbit which is less eccentric ($R_{apo} =
49.3$ kpc, $R_{peri} = 12.5$ kpc, $\epsilon = 0.60$) and in which the
cluster today is closer to apogalacticon. We will refer to this orbit
as \textit{orbit~2}. In addition, we search for the proper motion
values which minimize Pal~13's 3D orbital velocity, $V_{orb}$, since
this would yield the orbit in which Pal~13 is closest to
apogalacticon. In fact, by doing this we get a more eccentric orbit as
the cluster falls deeper into the Galactic centre ($R_{apo} = 38.5$
kpc, $R_{peri} = 3.5$ kpc, $\epsilon = 0.83$). This orbit will be
referred to as \textit{orbit 3}. It is also depicted in
Fig.~\ref{orbits}, with the corresponding values for the proper motion
being $\mu_\alpha\cos\delta = 0.72$ mas/yr and $\mu_\delta = -1.2$
mas/yr. Note that these values differ significantly from the
  values measured by \citet{Siegel01} by more than 1 mas/yr in each
  direction.

For all three orbits, the cluster is coming from perigalacticon and approaching apogalacticon. But all three predict Pal~13 to be in a different orbital phase, $p_{orb}$, which we here define as
\begin{equation}
p_{orb} = \frac{\dot{R}_{GC}}{|\dot{R}_{GC}|}\frac{R_{GC}-R_{peri}}{R_{apo}-R_{peri}},
\end{equation}
and which is constructed to be zero in perigalacticon and unity in apogalacticon. This factor $p_{orb}$ gives the fraction of the radial distance between $R_{peri}$ and $R_{apo}$ at which a cluster is currently located. In this definition $\dot{R}_{GC}$ is the time derivative of the galactocentric radius, which divided by its magnitude adds a minus sign to the orbital phase in case the cluster is moving from apogalacticon to perigalacticon. For circular orbits $p_{orb}$ is always zero.

We introduce the orbital phase, $p_{orb}$, here in addition to the orbital eccentricity, $\epsilon$, since it appears crucial for the appearance of the observed effects \citep{Kuepper10b}. A systematic study on the dependence of the described effects on those two parameters will follow in a future investigation. For \textit{orbit 1} we get $p_{orb} =0.20$, for \textit{orbit 2} $p_{orb} = 0.35$, and for \textit{orbit 3} $p_{orb} = 0.62$.

\subsection{Surface density profile}
\citet{Kuepper10b} introduce a template (KKBH) which can be used to reliably measure the slope of a surface density profile at large radii. The KKBH template reads as follows:
\begin{eqnarray}\label{eq:ETC1}
f_1(R) &=& k \left[\frac{R/R_c}{1+R/R_c}\right]^{-\gamma} \times\nonumber\\ 
& & \left[ \frac{1}{\sqrt{1+\left(R/R_c\right)^2}} - \frac{1}{\sqrt{1+\left(R_t/R_c\right)^2}} \right]^2
\end{eqnarray}
for radii smaller than $\mu\,R_t$, and
\begin{equation}\label{eq:ETC2}
f_2(R) = f_1(\mu\,R_t) \left[1 + \left(\frac{R}{\mu\,R_t}\right)^{64}\right]^{-\eta/64}
\end{equation}
for $R\ge\mu\,R_t$, where $k$ is a constant, $R_c$ gives a core radius, $\gamma$ the core slope inside $R_c$, and $R_t$ a tidal radius (which K\"upper et al.~name edge radius to avoid confusion with the theoretical tidal radius, since those two correlate only under certain circumstances). For radii larger than a fraction $\mu$ of $R_t$ the template changes into a power-law with slope $\eta$. The exponent 64 in $f_2(R)$ causes the template to change abruptly into the power-law slope at $\mu R_t$. 

KKBH is based on the template of \citet{King62} but is modified in two steps: first it allows to have a power-law cusp in the core, and second it has an additional extra-tidal component in the form of a power-law slope. In this way, also more concentrated clusters can be represented for which the original King template fails, and furthermore the cluster profile can be fitted without being influenced by a dominant tidal debris. This effect can be seen in Fig.~\ref{sdpCFHT} where we applied a regular King template fit to the CFHT data by C\^ot\'e et al.~and also a fit of KKBH\footnote{K\"upper et al.~recommend using an additional constant background, $b$, for the KKBH fit to allow for more flexibility at large radii, but as it turns out, this is only reasonable with highly resolved (e.g., $N$-body) data. For less well resolved observational data with much fewer data points, it is more reasonable to reduce the number of fit parameters to a minimum. We therefore set the background, $b$, mentioned in \citet{Kuepper10b} to zero, and in addition fix the break radius parameter $\mu$ to 0.5, as was found by K\"upper et al.~to be the most plausible value.}. The King template gets significantly influenced by the stellar material at large radii and yields a tidal radius of more than 15 arcmin. The KKBH template assigns a power-law slope of $\eta = 1.91 \pm 0.15$ to the tidal debris and yields a tidal radius of only 1.9 arcmin.\\ 

By computing $N$-body models for all three kinds of orbits given above, and fitting the KKBH template in the same way to similarly resolved $N$-body data, we will try to reproduce this observed surface density slope $\eta$. Furthermore we will measure the velocity dispersion and absolute magnitude of the computed clusters in the same way as C\^ot\'e et al.~have done, and compare these values with the observational ones.

\section{Models}\label{Sec:Models}
\begin{figure}
\includegraphics[width=84mm]{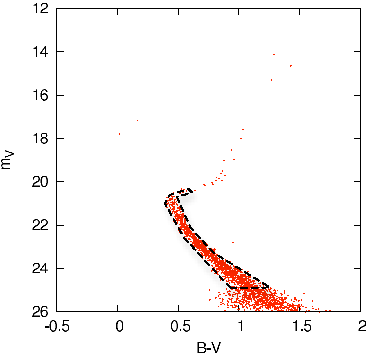}
  \caption{Mock colour-magnitude diagram of one of the computed clusters at the end of 3 Gyr of $N$-body integration, i.e. when the stellar population is 13 Gyr old. The y-axis gives the apparent V magnitude as would be observed from the distance of the Sun. Random errors which grow exponentially with increasing magnitude were applied to both apparent magnitudes, $m_V$ and $m_B$ (see text). This particular cluster had an initial mass of 5000 $\msun$ and a half-mass radius of 8 pc. The dashed box shows the region in the colour-magnitude diagram in which we define stars to be cluster members.}
  \label{CMD}
\end{figure}
\begin{figure}
\includegraphics[width=84mm]{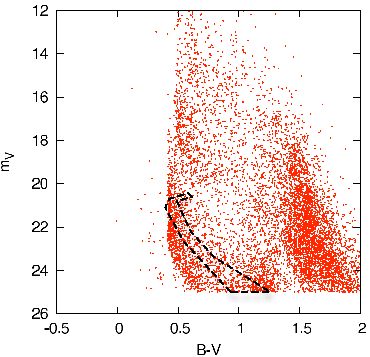}
  \caption{Colour-magnitude diagram of the Besan\c{c}on model \citep{Robin03} covering a 1 deg$^2$ field around the position of Pal~13. Within the region which is occupied by Pal~13 in this diagram (dashed box) we count about 1000 stars which will pollute observations when going down to $m_V = 25$ mag.}
  \label{besancon}
\end{figure}
We computed 45 models of Pal~13 using the collisional $N$-body code \textsc{NBODY6} \citep{Aarseth03} on the GPU computers at AIfA Bonn. We set up 15 different cluster configurations using the publicly available tool \textsc{McLuster}\footnote{\texttt{www.astro.uni-bonn.de/\~{}akuepper/mcluster/mcluster.html\\or  www.astro.uni-bonn.de/\~{}webaiub/german/downloads.php}} (K\"upper et al., in prep.). We used a tidally truncated Plummer profile where we varied the cluster half-mass radius between 4, 6 and 8 pc, and the initial mass between 3000, 4000, 5000, 7500 and 10000 $\msun$, respectively. The mean mass of the cluster stars was in all cases about $0.3\msun$, thus the number of objects in the computations were a factor of three times higher. Each of these 15 clusters was computed for the last 3 Gyr on each of the three different orbits (\textit{orbit~1-3}) mentioned in Sec.~\ref{Sec:Palomar 13} (see also Fig.~\ref{orbits}). We focus on the last 3 Gyr of evolution since we are only interested in the nearby tidal debris. Using equation 18 of \citet{Kuepper10a} for the mean drift velocity of stars within the tidal tails,
\begin{equation}
v_C = \pm (4GM\Omega)^{1/3},
\end{equation}
where $G$ is the gravitational constant, $M$ is the cluster mass and $\Omega$ its angular velocity on its orbit about the Milky Way, we can estimate the length of the tidal tails after 3 Gyr of evolution if the cluster was on a circular orbit. With the same values which we used in eq.~\ref{eq:Rt} we get a drift velocity of about 0.77 pc/Myr and thus a minimum length of the tails of 2.3 kpc in each direction from the cluster. Note that this estimate gets complicated through the fact that Pal~13 is most likely not on a circular orbit. Through the acceleration and deceleration on an eccentric orbit the cluster-tail system gets periodically stretched and compressed. This estimate is therefore only a mean value of the length of the tails. Anyway, for investigating the vicinity of Pal~13 this timespan seems to be sufficient.

Since we want to produce a realistic CMD of Pal 13 with the appropriate
photometric observables, we use the \textsc{SSE} code \citep{Hurley00} in combination with \textsc{McLuster}\footnote{Note that this version of \textsc{McLuster} including \textsc{SSE} is also available from the given web address.} to set up evolved stellar populations of 10 Gyr age with a metallicity of $[\mbox{Fe/H}] = -1.9$. The populations are evolved from a canonical Kroupa IMF \citep{Kroupa01} ranging from $0.08\msun$ to $100\msun$, where compact remnants are only kept if their kick velocity which gets assigned to them by \textsc{SSE} does not exceed Pal~13's present-day escape velocity, $v_{esc}$, calculated using
\begin{equation}
v_{esc} = \sqrt{\frac{2GM}{R_h}},
\end{equation}
where $M$ is again the cluster mass and $R_h$ is the cluster's half-mass radius, respectively. This treatment is a bit arbitrary since we do not dynamically model the first 10 Gyr of the cluster's life and the true retention fraction could be both, higher or lower. Another simple treatment would be to keep all compact remnants which would have a slight effect on the observed mass-to-light ratio as the cluster would then have more mass than can be seen in stars. But only about 80 compact remnants get expelled from a $5000\msun$ cluster like we model here in the way described above, hence we consider this to be of secondary importance and concentrate on the case of low dark mass in the clusters. 

Those evolved clusters we then feed to \textsc{NBODY6} to evolve them further, chemically and dynamically, up to a total age of 13 Gyr. In this way we can concentrate on the last few Gyr of dynamical evolution of the cluster, which are most important for its present-day structure and nearby tidal debris. Hence, we save computational time with this technique. A similar approach has been successfully tried by \citet{Hurley01} for the open cluster M67.

The stellar evolution of single stars within \textsc{NBODY6} is also calculated with \textsc{SSE}, a consistent treatment of stellar evolution throughout the investigation is therefore guaranteed. From \textsc{NBODY6} we finally extract the luminosities and stellar radii of all stars within the calculations to compute their effective temperatures and with this their colours and magnitudes in the Johnson-Cousins system \citep{Bessell90}. We use the algorithm described in \citet{Flower96} to first derive the bolometric correction, $BC$, and  the colour index, $B-V$, and with this the absolute magnitude in the $V$-band, $M_V$, and in the $B$-band, $M_B$. Together with the distance information of each star we can then derive the apparent magnitudes, $m_V$ and $m_B$, respectively, and can apply a realistic cut-off at a magnitude limit of, e.g., $m_V = 25$ mag as would be achieved by an 8m-class telescope in a few minutes of integration. 

Since any observation obeys statistical and instrumental uncertainties, we furthermore apply a Gaussian-distributed random error, d$m$ to each apparent magnitude, $m_V$ and $m_B$, which increases with decreasing brightness as
\begin{equation}
\mbox{d}m = \sqrt{0.02^2 + 0.07 \times 10.0^{0.4\times \left(m-25.0\right)}}.
\end{equation}
This gives a minimum error of 0.02 mag, and an additional uncertainty which is of the order of 0.07 mag at 25 mag and decreases with increasing brightness. A final, synthetic colour-magnitude diagram of one of the clusters is given in Fig.~\ref{CMD}. 

Objects with low surface densities such as tidal debris will be largely affected by background/foreground source contamination. We therefore have to estimate the number of stars which will pollute our mock observations. For this purpose we generate an artificial stellar population using the Besan\c{c}on model \citep{Robin03} for a 1 deg$^2$ field around the position of Pal~13 in the same filter set (Fig~\ref{besancon}). Within the CMD region occupied by the cluster model we count about 1000 stars which can be mistaken as cluster members, corresponding to 0.28 stars/arcmin$^2$. That is, Pal~13 and its tidal debris will only be visible in those places where its surface density exceeds this value. In addition to our estimate and for the sake of being conservative, we will also discuss our results using the somewhat higher background surface density of 0.69 stars/arcmin$^2$ found observationally by \citet{Cote02}.

\section{Results}\label{Sec:Results}
\begin{figure*}
\includegraphics[width=76mm]{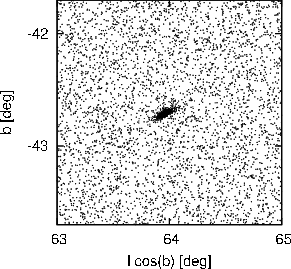}
\includegraphics[width=84mm]{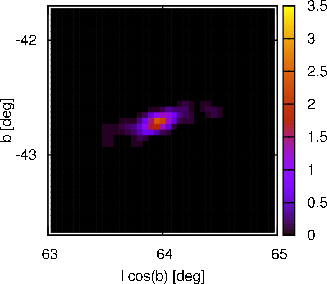}
  \caption{Left: stellar map of a 4 deg$^2$ field around Pal~13, computed using the orbit with the transverse velocity measured by \citet{Siegel01}, i.e. \textit{orbit~1}. Each dot represents a star above $m_V = 25$ mag. A background of 0.28 stars/arcmin$^2$ was added with random positions. Right: underlying surface density map of Pal~13 for the same field. One bin corresponds to 9 arcmin$^2$. The colour coding shows $\log_{10}(N+1)$, where $N$ is the number of Pal~13 stars in a bin. The expected background of 0.28 $[0.69]$ stars/arcmin$^2$ corresponds to a value of 0.5 $[0.9]$ in this representation. At the distance of Pal~13, 1 deg corresponds to roughly 420 pc. The cluster is well limited and the density falls off steeply, only small traces of tidal tails can be seen.}
  \label{map_v10}
\end{figure*}
\begin{figure*}
\includegraphics[width=76mm]{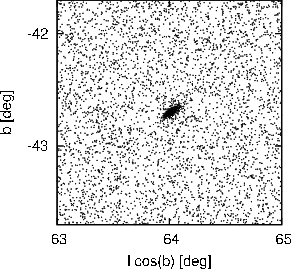}
\includegraphics[width=84mm]{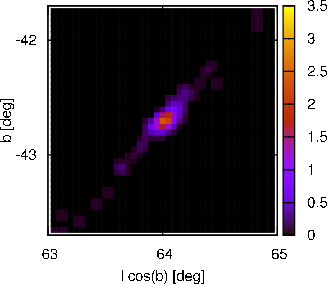}
  \caption{Left: stellar map of a 4 deg$^2$ field around Pal~13, computed using the orbit with zero transverse velocity (\textit{orbit~2}). Each dot represents a star above $m_V = 25$ mag. A background of 0.28 stars/arcmin$^2$ was added with random positions. Right: underlying surface density map of Pal~13 for the same field. One bin corresponds to 9 arcmin$^2$. The colour coding shows $\log_{10}(N+1)$, where $N$ is the number of Pal~13 stars in a bin. The expected background of 0.28 $[0.69]$ stars/arcmin$^2$ corresponds to a value of 0.5 $[0.9]$ in this representation. At the distance of Pal~13, 1 deg corresponds to roughly 420 pc. The cluster is also well limited, just as in Fig.~\ref{map_v10}. Only the tidal tails are a bit more pronounced since the cluster moves at a lower velocity and hence the stellar density within the tails is higher. }
  \label{map_v00}
\end{figure*}
\begin{figure*}
\includegraphics[width=76mm]{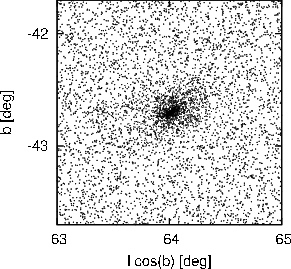}
\includegraphics[width=84mm]{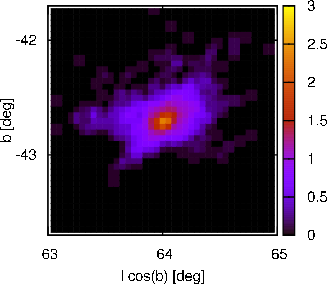}
  \caption{Left: stellar map of a 4 deg$^2$ field around Pal~13, computed using the orbit with the transverse velocity minimizing Pal~13's orbital velocity (\textit{orbit~3}). Each dot represents a star above $m_V = 25$ mag. A background of 0.28 stars/arcmin$^2$ was added with random positions. Right: underlying surface density map of Pal~13 for the same field. One bin corresponds to 9 arcmin$^2$. The colour coding shows $\log_{10}(N+1)$, where $N$ is the number of Pal~13 stars in a bin. The expected background of 0.28 $[0.69]$ stars/arcmin$^2$ corresponds to a value of 0.5 $[0.9]$ in this representation. At the distance of Pal~13, 1 deg corresponds to roughly 420 pc. The cluster is embedded in a far-extending cloud of stars, which originates from the compressed tidal tails getting pushed back into the cluster vicinity as the cluster-tail system is being decelerated on its way to apogalacticon.}
  \label{map_min}
\end{figure*}
From each computation we take a snapshot after 3 Gyr of dynamical evolution as seen from the location of the Sun. At this point the stellar population is 13 Gyr old, and should resemble the stellar population of Pal~13 within the given uncertainties. Stellar maps of a 4 deg$^2$ region around one of the clusters ($M_0 = 5000, R_0 = 8$ pc) for each of the three orbits are shown in  the left panels of Fig.~\ref{map_v10}-\ref{map_min}. In these figures each dot represents a star above $m_V = 25$ mag. A background of 0.28 stars/arcmin$^2$ was added with random positions. By comparing the figures we see that the orbit with the Siegel et al.~proper motion (\textit{orbit~1}, Fig.~\ref{map_v10}) and the one with zero proper motion (\textit{orbit~2}, Fig.~\ref{map_v00}) produce similar results, whereas the orbit with the minimal orbital velocity (\textit{orbit~3}, Fig.~\ref{map_min}) produces a  cluster which appears largely extended.

\subsection{Absolute magnitude}
\begin{table}
\begin{minipage}{84mm}
\centering
 \caption{Results for the $N$-body computations with the orbit using the transverse velocity measured by \citet{Siegel01}, i.e. \textit{orbit 1}. $M_0$ gives the initial mass of the cluster at the beginning of the computations, and $R_0$ its initial half-mass radius. $M_V$ is the measured absolute magnitude within the inner 26 arcmin at an age of 13 Gyr, i.e. today. \Citet{Cote02} find a value of $M_V = -3.8$ mag for Pal~13. $\sigma_r$ gives the velocity dispersion within the inner 2 arcmin measured from a sample of 21 stars. The value gives the mean of $10^6$ independent measurements (see Sec.~\ref{Sec:Velocity dispersion} for details), the uncertainties give the limits in which 67\% of all measurements lie. \citet{Cote02} find $\sigma_r = 2.2 \pm 0.4$ for Pal~13. $\eta$ is the slope of the surface density profile at large radii measured with the KKBH template. The uncertainties give the standard error from a least square fit. For Pal~13 we measure a slope of about 1.9 based on the observational data by Cote et al. (2002).}
\label{table1}
\begin{tabular}{ccccc}
\hline 
$M_0$ & $R_0$ & $M_V \, (R<26')$ & $\sigma_r \, (R<2')$ &  $\eta$ \\ 

[$\msun$] & [pc] &  [mag] &  [km/s] &  \\ 
\hline 
3000 & 4.0 & -2.7 & $0.53^{+0.09}_{-0.09}$ & 3.86$\pm$0.58 \\ 
3000 & 6.0 & -2.6 & $0.54^{+0.09}_{-0.09}$ & 3.55$\pm$0.34 \\ 
3000 & 8.0 & -2.1 & $0.46^{+0.07}_{-0.07}$ & 4.34$\pm$0.76 \\ 
4000 & 4.0 & -3.0 & $0.63^{+0.11}_{-0.11}$ & 3.33$\pm$0.20 \\ 
4000 & 6.0 & -3.2 & $0.63^{+0.11}_{-0.10}$ & 3.77$\pm$0.12 \\ 
4000 & 8.0 & -2.6 & $0.53^{+0.08}_{-0.08}$ & 4.15$\pm$0.16 \\ 
5000 & 4.0 & -4.0 & $0.69^{+0.12}_{-0.11}$ & 3.58$\pm$0.14 \\ 
5000 & 6.0 & -3.1 & $0.71^{+0.12}_{-0.11}$ & 3.82$\pm$0.15 \\ 
5000 & 8.0 & -4.0 & $0.58^{+0.09}_{-0.09}$ & 3.86$\pm$0.14 \\ 
7500 & 4.0 & -3.9 & $0.92^{+0.16}_{-0.16}$ & 3.36$\pm$0.07 \\ 
7500 & 6.0 & -4.2 & $0.84^{+0.14}_{-0.13}$ & 4.00$\pm$0.16 \\ 
7500 & 8.0 & -3.5 & $0.73^{+0.12}_{-0.11}$ & 4.08$\pm$0.07 \\ 
10000 & 4.0 & -3.7 & $1.11^{+0.20}_{-0.19}$ & 3.86$\pm$0.24 \\ 
10000 & 6.0 & -4.2 & $0.96^{+0.16}_{-0.15}$ & 4.40$\pm$0.21 \\ 
10000 & 8.0 & -3.6 & $0.83^{+0.13}_{-0.13}$ & 4.24$\pm$0.15 \\ 
\end{tabular}
\end{minipage}
\end{table}
\begin{table}
\begin{minipage}{84mm}
\centering
 \caption{The same as Tab.~\ref{table1} but for the orbit with zero transverse velocity (\textit{orbit~2}).}
\label{table2}
\begin{tabular}{ccccc}
\hline 
$M_0$ & $R_0$ & $M_V \, (R<26')$ & $\sigma_r \, (R<2')$ &  $\eta$ \\ 

[$\msun$] & [pc] &  [mag] &  [km/s] &  \\ 
\hline 
3000 & 4.0 & -2.7 & $0.52^{+0.09}_{-0.09}$ & 3.10$\pm$0.41 \\ 
3000 & 6.0 & -2.5 & $0.52^{+0.09}_{-0.09}$ & 4.04$\pm$0.13 \\ 
3000 & 8.0 & -2.1 & $0.44^{+0.07}_{-0.07}$ & 3.47$\pm$0.28 \\ 
4000 & 4.0 & -3.0 & $0.64^{+0.11}_{-0.11}$ & 3.35$\pm$0.23 \\ 
4000 & 6.0 & -3.1 & $0.63^{+0.11}_{-0.11}$ & 3.98$\pm$0.12 \\ 
4000 & 8.0 & -2.6 & $0.51^{+0.08}_{-0.08}$ & 4.07$\pm$0.40 \\ 
5000 & 4.0 & -4.1 & $0.72^{+0.12}_{-0.12}$ & 3.47$\pm$0.07 \\ 
5000 & 6.0 & -3.1 & $0.69^{+0.11}_{-0.11}$ & 3.59$\pm$0.33 \\ 
5000 & 8.0 & -4.0 & $0.56^{+0.09}_{-0.09}$ & 4.05$\pm$0.28 \\ 
7500 & 4.0 & -3.9 & $0.92^{+0.16}_{-0.16}$ & 3.53$\pm$0.14 \\ 
7500 & 6.0 & -4.2 & $0.83^{+0.13}_{-0.13}$ & 4.20$\pm$0.29 \\ 
7500 & 8.0 & -3.5 & $0.71^{+0.11}_{-0.11}$ & 4.48$\pm$0.24 \\ 
10000 & 4.0 & -3.7 & $1.08^{+0.19}_{-0.19}$ & 3.82$\pm$0.29 \\ 
10000 & 6.0 & -4.2 & $0.96^{+0.15}_{-0.15}$ & 4.05$\pm$0.22 \\ 
10000 & 8.0 & -3.6 & $0.82^{+0.12}_{-0.13}$ & 4.47$\pm$0.22 \\ 
\end{tabular}
\end{minipage}
\end{table}
\begin{table}
\begin{minipage}{84mm}
\centering
 \caption{The same as Tab.~\ref{table1} but for the orbit with the transverse velocity minimizing Pal~13's orbital velocity (\textit{orbit~3}). Velocity dispersion values which agree within 1$\sigma$ with the observed velocity dispersion of $2.2\pm0.4$ are bold faced.}
\label{table3}
\begin{tabular}{ccccc}
\hline 
$M_0$ & $R_0$ & $M_V \, (R<26')$ & $\sigma_r \, (R<2')$ &  $\eta$ \\ 

[$\msun$] & [pc] &  [mag] &  [km/s] &  \\ 
\hline 
3000 & 4.0 & -2.7 & $\textbf{0.92}^{+1.13}_{-0.48}$ & 1.90$\pm$0.07 \\ 
3000 & 6.0 & -2.6 & $\textbf{1.21}^{+0.96}_{-0.75}$ & 1.71$\pm$0.14 \\ 
3000 & 8.0 & -2.1 & $\textbf{1.69}^{+0.93}_{-1.24}$ & 1.48$\pm$0.07 \\ 
4000 & 4.0 & -3.0 & $\textbf{0.73}^{+1.06}_{-0.20}$ & 2.13$\pm$0.09 \\ 
4000 & 6.0 & -3.2 & $\textbf{0.99}^{+1.14}_{-0.47}$ & 1.68$\pm$0.10 \\ 
4000 & 8.0 & -2.6 & $\textbf{1.15}^{+1.10}_{-0.76}$ & 1.51$\pm$0.09 \\ 
5000 & 4.0 & -4.1 & $0.75^{+0.17}_{-0.17}$ & 1.90$\pm$0.10 \\ 
5000 & 6.0 & -3.1 & $0.80^{+0.39}_{-0.20}$ & 1.51$\pm$0.10 \\ 
5000 & 8.0 & -4.0 & $\textbf{1.11}^{+1.14}_{-0.63}$ & 1.62$\pm$0.10 \\ 
7500 & 4.0 & -3.9 & $0.94^{+0.16}_{-0.17}$ & 2.02$\pm$0.15 \\ 
7500 & 6.0 & -4.2 & $0.90^{+0.18}_{-0.19}$ & 2.16$\pm$0.20 \\ 
7500 & 8.0 & -3.5 & $\textbf{0.89}^{+1.06}_{-0.28}$ & 1.59$\pm$0.21 \\ 
10000 & 4.0 & -3.7 & $1.16^{+0.20}_{-0.20}$ & 2.52$\pm$0.11 \\ 
10000 & 6.0 & -4.2 & $0.96^{+0.16}_{-0.17}$ & 2.66$\pm$0.30 \\ 
10000 & 8.0 & -3.6 & $0.87^{+0.56}_{-0.19}$ & 1.96$\pm$0.31 \\ 
\end{tabular}
\end{minipage}
\end{table}
From these snapshots we measure the integrated absolute magnitude, $M_V$, of each cluster representation within a radius of 26 arcmin around its centre, just as \citet{Cote02} have done for their observational data. The results are listed in Tab.~\ref{table1}-\ref{table3} for the three orbital types. We see that the clusters starting off with smaller initial masses, independent of the orbit, have lost too much mass within the 3 Gyr of evolution, such that today their absolute magnitude is too low compared to the observational value of $M_V = -3.8$ mag. Clusters with $M_0 \ge 5000 \msun$ lose just about the right amount of mass within this time. From the tables we can expect clusters with initial masses even higher than $10000 \msun$ to exceed the observed absolute magnitude. This suggests that our range of initial parameters covers the right part of the parameter space of initial conditions.

Moreover, comparing the same clusters but on the different orbits, we find that the clusters evolve quite similarly   internally and that their absolute magnitudes are only marginally influenced by the orbital type. In fact, at the beginning of the computations the clusters of a given mass and size are exactly the same clusters just on different orbits. In this way we make sure that differences come from dynamical evolution and not from stellar evolution. After 3 Gyr the masses between the clusters of a given initial mass and size differ by only $50-100\msun$. From this we can deduce that the influence of the pericentre passages on all three orbits are rather unimportant. Otherwise, the more eccentric orbits (\textit{orbit~3} and \textit{orbit~1}) would have induced more dissolution on those clusters, and altered the final absolute magnitudes more significantly.

\subsection{Velocity dispersion}\label{Sec:Velocity dispersion}
From the computations we also take radial velocity dispersion measurements in the same way as \citet{Cote02} have done. That is, we draw 21 stars from the sample of stars within the inner 2 arcmin of the clusters, while making sure that all 21 stars lie within 10 km/s of each other. A star with a radial velocity differing more than 10 km/s from the other stars which have been drawn from the population would therefore be regarded as a non-cluster member, even though this does not necessarily hold true in our computations. The velocity dispersion is then computed in the same fashion as has been done in \citet{Kuepper10c}. We independently draw $10^6$ sets of 21 stars from each cluster and compute for each set the dispersion of the stellar velocities. From these values we take the mean, which is given in Tab.~\ref{table1}-\ref{table3}. The uncertainties of these values give the bounds in which lie 67\% ($1\sigma$) of all measurements above and below the mean. 

From Tab.\ref{table1} we can see that the clusters on the Siegel orbit (\textit{orbit~1}) yield too low velocity dispersions in comparison to the observational value of $\sigma_r = 2.2 \pm 0.4$ km/s. The orbit with zero proper motion yields similar results (\textit{orbit~2}, Tab.~\ref{table2}). In both sets of computations we achieve the highest velocity dispersions of $1.1\pm0.2$ km/s in the most massive and most compact cluster of  $10000 \msun$ and $R_0 = 4.0$ pc. This is expected when we assume that the clusters are in virial equilibrium. Note that the observed velocity dispersion is more than $5 \sigma$ off. This fact led C\^ot\'e et al.~to the assumption that Pal~13 may contain dark matter or got catastrophically heated by the last pericentre passage.

But for the third kind of orbit, when the cluster is on an orbit with a lower orbital velocity such that it is nowadays closer to apogalacticon (\textit{orbit~3}), the measured velocity dispersion is significantly higher (Tab.~\ref{table3}). We get values of about 2.2 km/s within $1\sigma$ for many clusters in the set. This is due to the number of unbound stars within the cluster, so-called potential escapers, and  stars outside the tidal radius lying in projection within the inner 2 arcmin of the cluster, which pollute the velocity dispersion \citep{Kuepper10b}. This effect is more significant for the clusters which are initially more extended as they show more potential escapers. This is a consequence of them being energetically more affected by the pericentre passages (see e.g. \citealt{Gnedin99}).

The clusters on \textit{orbit~3} indeed show much more extra-tidal material. Looking at the right panels of Fig.~\ref{map_v10}-\ref{map_min} we see that there is barely any stellar material outside the cluster for \textit{orbit~1} (Fig.~\ref{map_v10}), and only little more for \textit{orbit~2} (Fig.~\ref{map_v00}). In contrast to this, \textit{orbit~3} shows a cluster with an unusual extra-tidal extent of several hundred parsec (Fig.~\ref{map_min}). This extra-tidal material results from the compression of the tidal tails as the cluster and its tails are being decelerated. This deceleration is so strong that the whole system consisting of cluster, leading tail, and trailing tail, is compressed to a few hundred pc. In fact, the system extends even further than can be seen in the figure and would extend much further if the $N$-body computations would have been made for the full 13 Gyr since the tidal tails need many Gyr to grow to such extent. Note also, that the shape of this system is quite irregular since we look at a folded stellar stream and not at a bound structure in equilibrium.

Note that while our choice of the initial density profile may well affect the measured velocity dispersion, e.g. in case of a higher concentrated King model, it will not affect the appearance of the tidal debris. That is, the choice of profile may influence the internal structure of the cluster and also its mass loss rate but not its tidal debris since the debris is formed by orbital compression and not by the mass loss rate \citep{Kuepper10b}.

\subsection{Surface density profiles}
\begin{figure*}
\includegraphics[width=55mm]{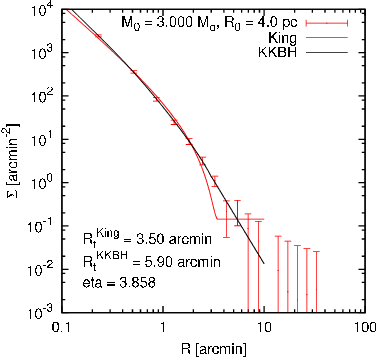}
\includegraphics[width=55mm]{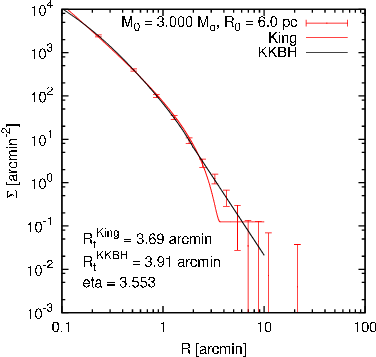}
\includegraphics[width=55mm]{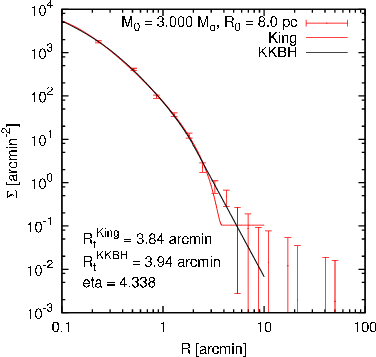}\\
\includegraphics[width=55mm]{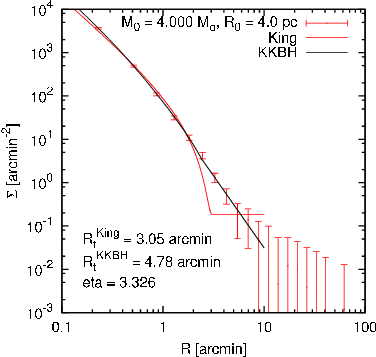}
\includegraphics[width=55mm]{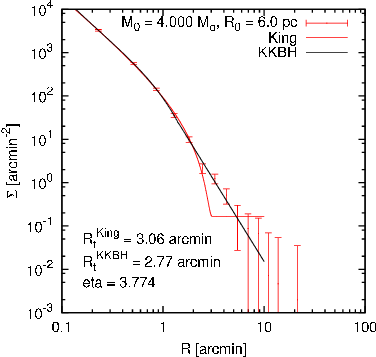}
\includegraphics[width=55mm]{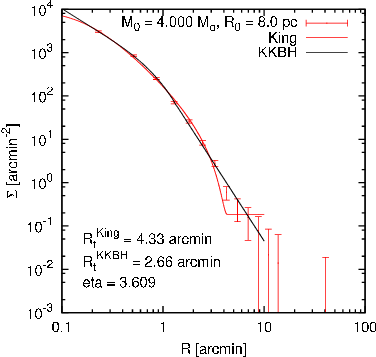}\\
\includegraphics[width=55mm]{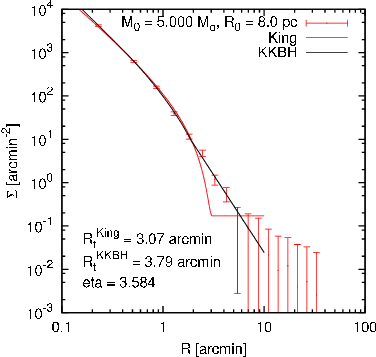}
\includegraphics[width=55mm]{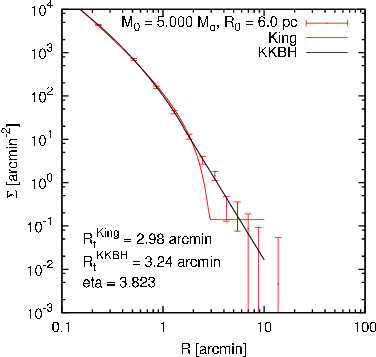}
\includegraphics[width=55mm]{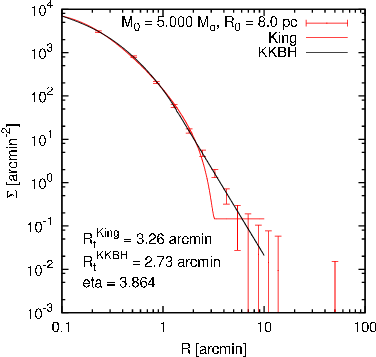}\\
\includegraphics[width=55mm]{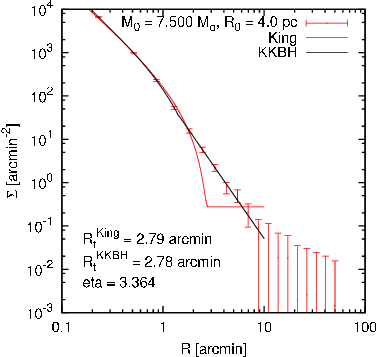}
\includegraphics[width=55mm]{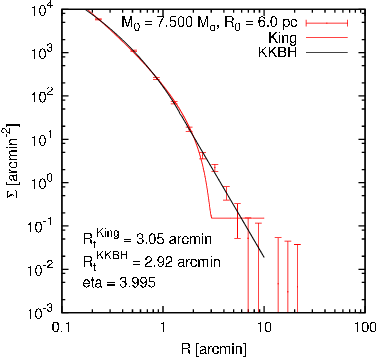}
\includegraphics[width=55mm]{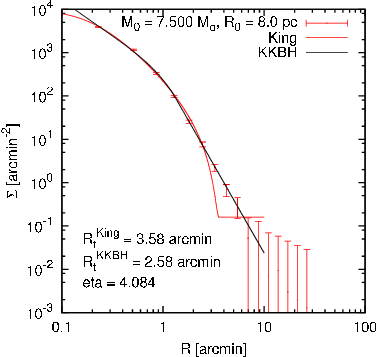}\\
  \caption{Surface density profiles of the $N$-body computations with the transverse velocity measured by \citet{Siegel01}, i.e. \textit{orbit~1}, for all clusters between $3000\msun$ and $7500\msun$. Uncertainties show the square-root of the number of stars in one bin after subtracting a background of 0.69 stars/arcmin$^2$ like \citet{Cote02} have done (compare with Fig.~\ref{sdpCFHT}). The slopes at large radii as measured with the KKBH template are given in the panels ($\eta$, i.e., $eta$). Also given in the panels are the tidal radii as fitted by the King and the KKBH template. All slopes at large radii are quite steep, just as would be expected for a cluster near perigalacticon \citep{Kuepper10b}.}
  \label{sdpv10}
\end{figure*}
\begin{figure*}
\includegraphics[width=55mm]{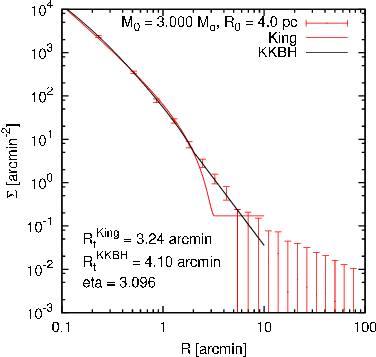}
\includegraphics[width=55mm]{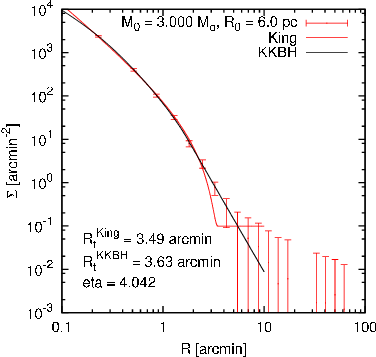}
\includegraphics[width=55mm]{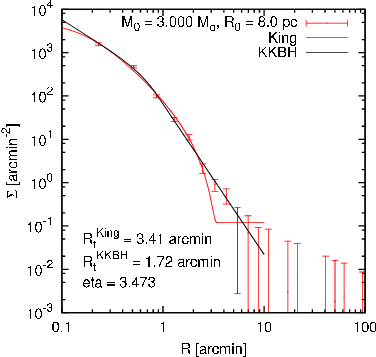}\\
\includegraphics[width=55mm]{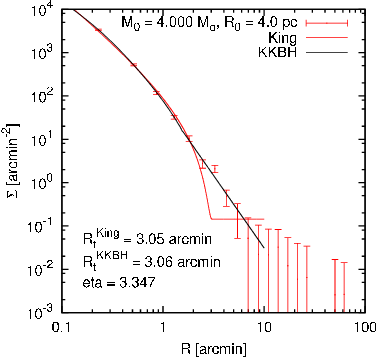}
\includegraphics[width=55mm]{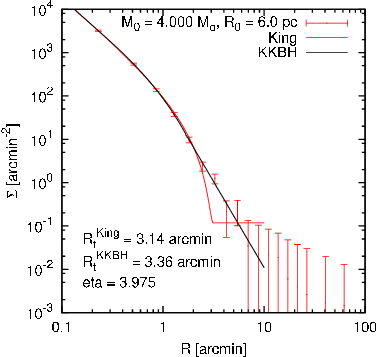}
\includegraphics[width=55mm]{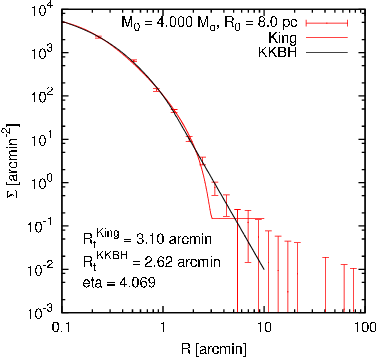}\\
\includegraphics[width=55mm]{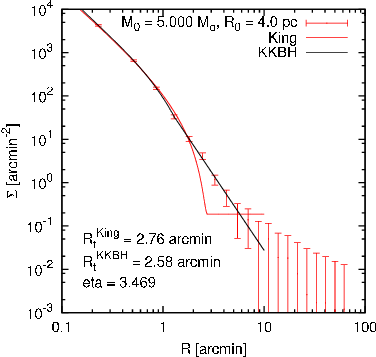}
\includegraphics[width=55mm]{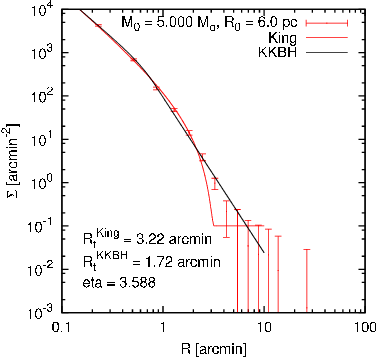}
\includegraphics[width=55mm]{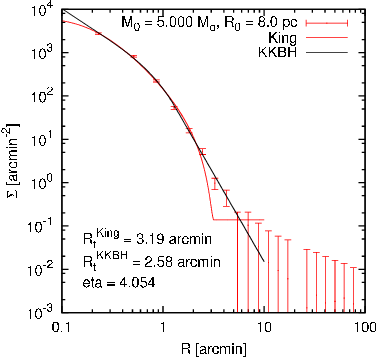}\\
\includegraphics[width=55mm]{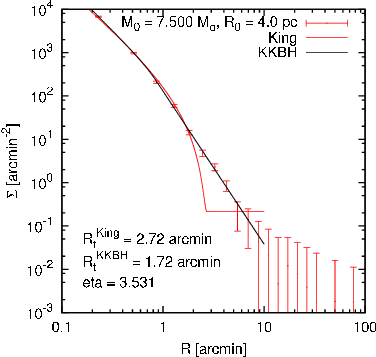}
\includegraphics[width=55mm]{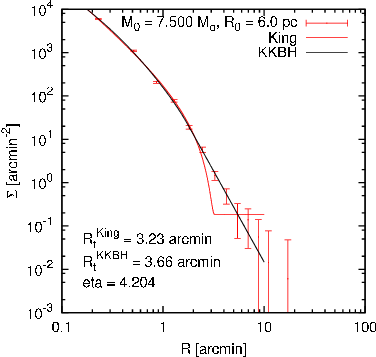}
\includegraphics[width=55mm]{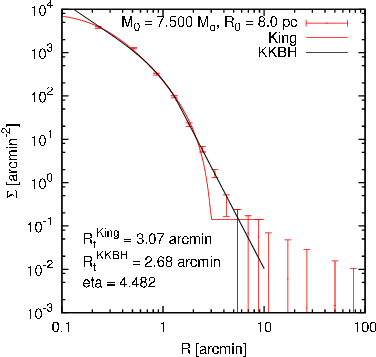}\\
  \caption{Surface density profiles of the $N$-body computations with zero transverse velocity (\textit{orbit~2} for all clusters between $3000\msun$ and $7500\msun$. Uncertainties show the square-root of the number of stars in one bin after subtracting a background of 0.69 stars/arcmin$^2$ like \citet{Cote02} have done (compare with Fig.~\ref{sdpCFHT}). The slopes at large radii as measured with the KKBH template are given in the panels ($\eta$, i.e., $eta$). Also given in the panels are the tidal radii as fitted by the King and the KKBH template. All slopes at large radii are quite steep, just as would be expected for a cluster near perigalacticon \citep{Kuepper10b}.}
  \label{sdpv00}
\end{figure*}
\begin{figure*}
\includegraphics[width=55mm]{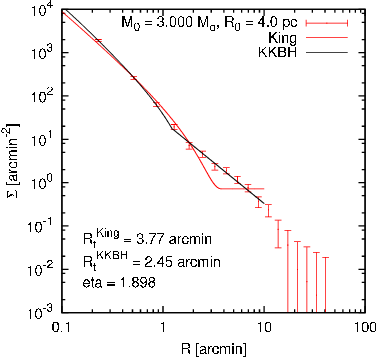}
\includegraphics[width=55mm]{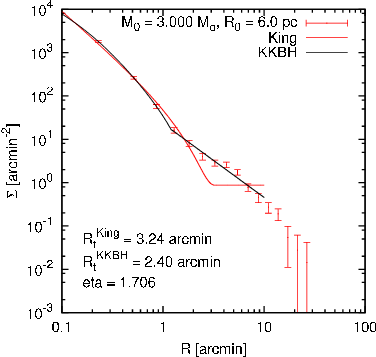}
\includegraphics[width=55mm]{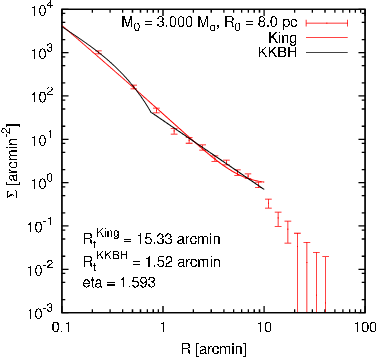}\\
\includegraphics[width=55mm]{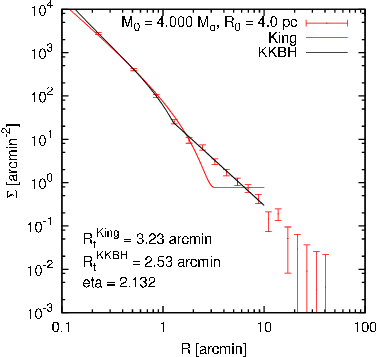}
\includegraphics[width=55mm]{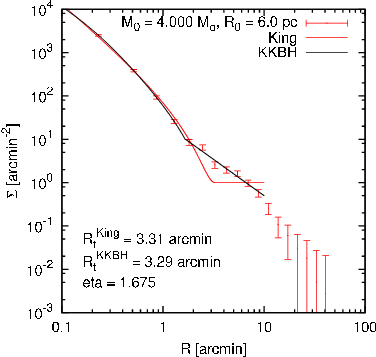}
\includegraphics[width=55mm]{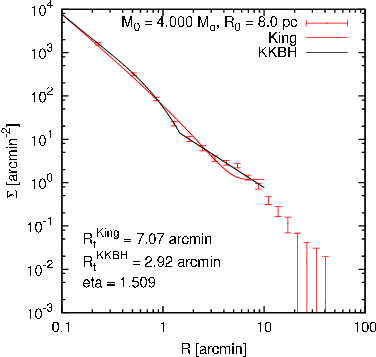}\\
\includegraphics[width=55mm]{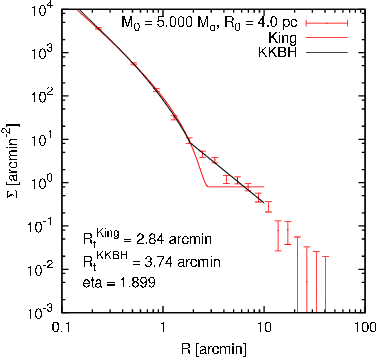}
\includegraphics[width=55mm]{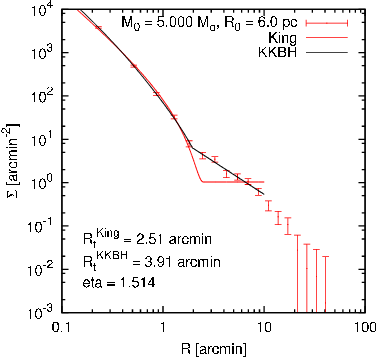}
\includegraphics[width=55mm]{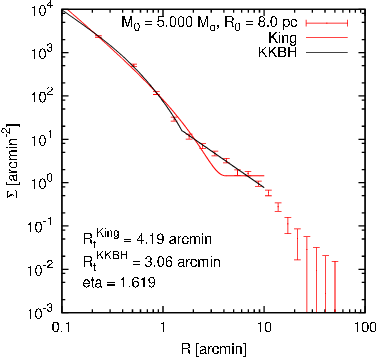}\\
\includegraphics[width=55mm]{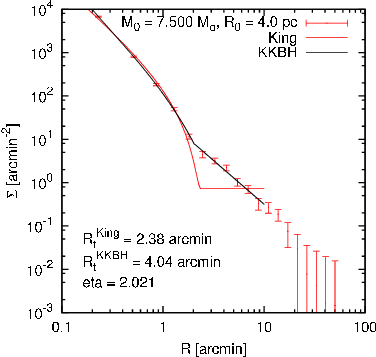}
\includegraphics[width=55mm]{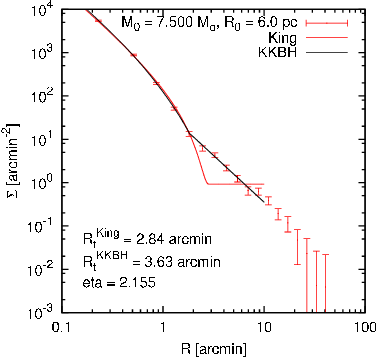}
\includegraphics[width=55mm]{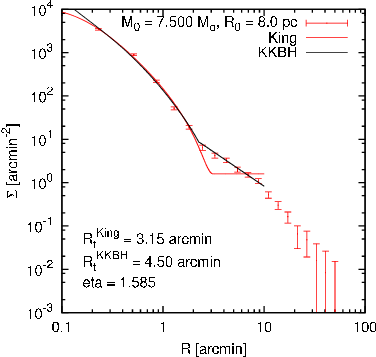}\\
  \caption{Surface density profiles of the $N$-body computations with the transverse velocity minimizing Pal~13's orbital velocity (\textit{orbit~3}) for all clusters between $3000\msun$ and $7500\msun$. Uncertainties show the square-root of the number of stars in one bin after subtracting a background of 0.69 stars/arcmin$^2$ like \citet{Cote02} have done (compare with Fig.~\ref{sdpCFHT}). The slopes at large radii as measured with the KKBH template are given in the panels ($\eta$, i.e., $eta$). Also given in the panels are the tidal radii as fitted by the King and the KKBH template. In contrast to the other two orbits, this orbit yields surface density profiles with shallow slopes at large radii. This comes from the deceleration of the cluster-tail system on its way to apogalacticon and the compression it causes. }
  \label{sdpvmin}
\end{figure*}
This effect of compression of the tidal debris can also be seen as an increase of density within the surface density profiles of the modelled clusters, see Figs.~\ref{sdpv10}-\ref{sdpvmin}. The figures show the projected stellar number density of the snapshots for all clusters except for the most massive ones, measured in rings around the cluster centres, just as \citet{Cote02} have done it with their CFHT data (compare with Fig.~\ref{sdpCFHT}). For better comparison with the CFHT data, we subtract the same background of 0.69 stars/arcmin$^2$ from our data instead of our lower estimate of 0.28 stars/arcmin$^2$ since this background estimate is important for the outermost data points and therefore may influence the fit of the KKBH template. Error bars in the figures give the square-root of these resulting values as statistical uncertainties. Differences in the numbers of stars at small radii within this diagram mainly originate from the fact that Cote et al.~go down to $m_V = 23.5$ mag with their CFHT data whereas we cut at $m_V = 25$ mag (assuming that Pal13 is observed with an 8m class telescope). We fit a KKBH template to these surface density profiles in order to measure the slope of the extra-tidal material. The results of these fits are displayed in the figures ($\eta$, i.e., $eta$) as well as in Tab.~\ref{table1}-\ref{table3}.

The Siegel orbit (\textit{orbit~1}) yields well limited clusters with steep slopes outside the tidal radius between $\eta = 3.3$ and $\eta = 4.4$ (Fig.~\ref{sdpv10}), just as expected from clusters near perigalacticon \citep{Kuepper10b}. Furthermore, there is no clear trend in the slopes with respect to the initial conditions. The differences are just the statistical fluctuations which K\"upper et al.~also found in their $N$-body data. The same holds true for \textit{orbit~2} (Fig.~\ref{sdpv00}). Just the scatter is a bit larger from $\eta = 3.1$ to $\eta =  4.5$ but, again, without any clear trend. If the slope at large radii was a consequence of the last pericentre passage, then we would expect a correlation of this slope with the initial half-mass radius, $R_0$, of the cluster, since a more extended cluster should be more affected by tidal shocking and therefore produce a more pronounced tidal debris.

In contrast to that, \textit{orbit~3} results in quite different surface density profiles (Fig.~\ref{sdpvmin}). We get a shallow slope at large radii for all clusters with values as low as $\eta = 1.5$, and for the steepest not more than $\eta  = 2.7$. That is, the clusters near apogalacticon differ significantly from the clusters which are closer to perigalacticon. Such a behaviour of the surface density profile has been found in observations of other globular clusters as well. For instance, Palomar~5, which is known to be close to its apogalacticon, shows an $\eta$ of 1.5 \citep{Odenkirchen03}. \citet{Chun10} find 5 Milky Way globular clusters to show shallow values of $\eta$ at large radii, that is 2.44 for NGC 5466, 1.59  for M 15, 1.58  for M 53, 1.41 for M 30, and 0.62  for NGC 5053. Moreover, AM 4 and Whiting 1 both show an $\eta$ of 1.8 \citep{Carraro07, Carraro09}. \citet{Olszewski09} furthermore find the Galactic globular cluster NGC 1851 to be surrounded by a 500 pc halo of stars. Its surface density profile shows a slope of $\eta = 1.24\pm0.66$. Our investigation suggests that those clusters are all affected by this orbital effect. 

From Fig.~\ref{sdpvmin} we furthermore find that fitting a \citet{King62} template to these clusters with the shallow slopes at large radii sometimes yields very different results for the tidal radius in comparison to the KKBH template (e.g. for $M_0 = 3000 \msun$ and $R_0 = 8$ pc we get $R_t^{KKBH} = 1.5$ arcmin and $R_t^{King} = 15.3$ arcmin). This we have observed as well in the original data (Fig.~\ref{sdpCFHT}) and thus would be an explanation for the large uncertainties in Pal~13's tidal radius \citep{Siegel01, Cote02}.

\section{Discussion}\label{sec:discussion}
\begin{figure*}
\includegraphics[width=84mm]{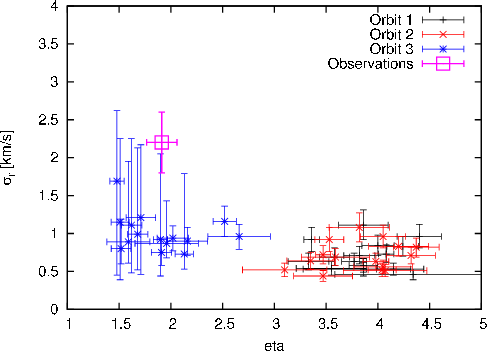}
\includegraphics[width=84mm]{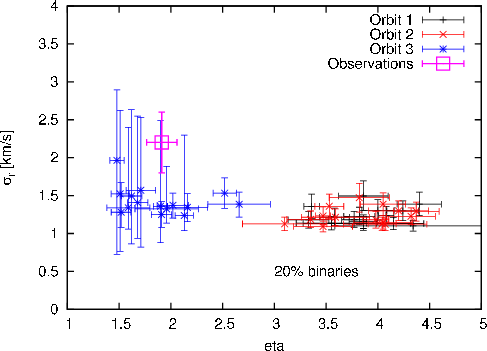}\\
\includegraphics[width=84mm]{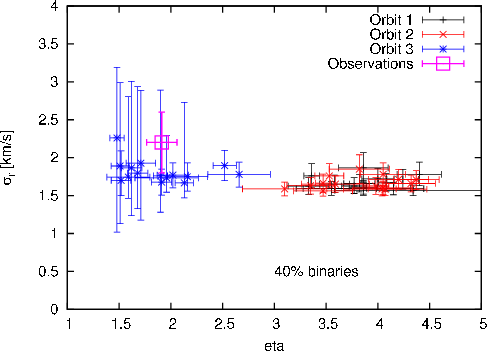}
\includegraphics[width=84mm]{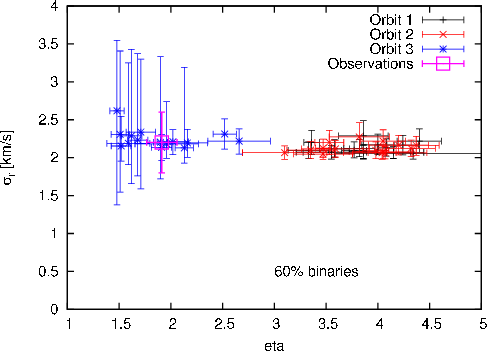}
  \caption{Upper left panel: overview of the computation results showing the $\eta$ ($eta$) values measured within the surface density profiles, and the measured radial velocity dispersions, $\sigma_r$, of all models in comparison to values observed by \citet{Cote02}. Each model is an independent $N$-body realisation of Pal~13 evaluated at an age of 13 Gyr (see Sec.~\ref{Sec:Models}). Other panels: expected velocity dispersions after correcting the measurements for 20\% (upper right), 40\% (lower left) and 60\% (lower right) binaries. Adding about 20-40\% binaries improves the match between the results of \textit{orbit 3} and the observations, whereas \textit{orbit 1} and \textit{orbit 2} necessitate $>$40\% binaries in order to agree well with the observed velocity dispersion.}
  \label{results}
\end{figure*}
\begin{figure}
\includegraphics[width=84mm]{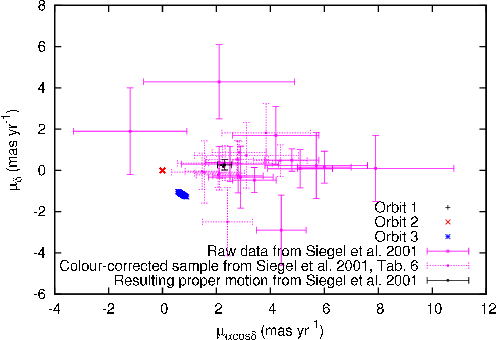}
  \caption{Comparison of proper motion measurement from our computations and from the work of \citet{Siegel01}. The data points of \textit{orbit 1 - 3} show the proper motions of all stars within the inner 6 arcmin of the clusters. As expected, the data points of \textit{orbit 1} all match precisely with the orbit determined by Siegel et al., while the data points of \textit{orbit 2} all lie concentrated within the origin at zero proper motion. The stars of the cluster on \textit{orbit 3} show a significant intrinsic spread, though, which is due to stars belonging to the cluster's tidal debris and which therefore do not tightly follow the bulk cluster motion. The data from Siegel et al.~is shown with and without colour correction to illustrate the large spread and uncertainties of the raw data. Also shown is the resulting proper motion found by this group.}
  \label{propermotion}
\end{figure}
Judging by the different measurements we have performed on our
artificial clusters (upper left panel of Fig.~\ref{results}) we see that the cluster starting off with $M_0 =
5000 \msun$ and $R_0 = 8$ pc on the orbit with the minimal orbital
velocity agrees best with the observational values found by
\citet{Cote02}. The match is not perfect, though, and we do not
conclude from our findings that Pal~13 has to be on this specific
orbit with these specific values for mass and radius used here. We rather demonstrate that most evidence points to Pal~13 being in an orbital phase near apogalacticon (in disagreement to the proper motion estimates by \citealt{Siegel01}). Assuming this solves the existing problems of Pal~13 without further need for dark matter, binaries or changes in
the law of gravity. 

Nevertheless, although our results reproduce the observations of Pal~13 without further ado, we have to check how the computational results change when adding binaries to the clusters. That is, Pal~13 most likely has a significant binary population, which we have not modelled in our $N$-body calculations, but which should inflate the velocity dispersion further. In fact, \citet{Blecha04} measure significantly different radial velocities for some of the C\^ot\'e et al.~sources with independent FLAMES data. This might well be due to binary motion or, alternatively, to a different quality of the instruments used. From 7 stars which are available in both studies 2--3 show significant variations, hinting at a high binary fraction. Thus, Blecha et al.~get velocity dispersion values between 0.6 km/s and 0.9 km/s from samples of 5--8 stars.

\citet{Kuepper10c} show that for the outer halo MW globular cluster Palomar~14 the mean velocity dispersion measured from samples of 17 stars increases from about 0.5 km/s in the case of Pal~14 being without binaries up to almost 4 km/s in the case of it having 100\% binaries. Thus, adding [20, 40, 60]\% binaries to the cluster (i.e. out of 100 systems in the cluster [20, 40, 60] are binaries) increases the velocity dispersion by about [1.0, 1.5, 2.0] km/s when measured in this way. Since both, Pal~14 and Pal~13, are quite comparable in their expected virial-equilibrium velocity dispersion, the effect of their binary content on observations should be comparable, too. In Fig.~\ref{results} we therefore additionally show the results from our investigation corrected for a binary fraction of 20\%, 40\% and 60\%. The corrections, d$\sigma$, of 1.0, 1.5 and 2.0 km/s, respectively, were added to the measured velocity dispersions quadratically, i.e.,
\begin{equation}
\sigma_{corr} = \sqrt{\sigma_r^2+\mbox{d}\sigma^2}.
\end{equation}
As can be seen in the figure, a binary fraction of 20-40\% improves the match of our models on \textit{orbit 3} with the observed velocity dispersion, which is a reasonable binary fraction for a globular cluster \citep{Hut92}, whereas the models on \textit{orbit 1} and \textit{orbit 2} would prefer values of $>$40\% in order to agree with the observational value of the velocity dispersion. This would have no effect on the discrepant slope of the surface density profile at large radii of the models on \textit{orbit 1 \& 2}, though.

But why do our proper motion values of \textit{orbit 3} differ so significantly from the Siegel et al.~values? First of all, as stated above, we do not argue that Pal~13 has to be on this specific orbit but is most likely close to its apogalacticon. A family of proper motion values around the values we chose may reproduce the observations equally well, since we only chose those values in order to minimize Pal~13's orbital velocity within the MW such that the orbital compression effect is maximal. Moreover, the observational values of Siegel et al.~may also be influenced by the orbital compression effect just as the other measurements. Consequently, their uncertainties may be largely underestimated.

In Fig.~\ref{propermotion} we show the original raw data from \citet{Siegel01} which was obtained from photographic plates separated by a 40 year baseline. After determining the zero point of each photographic plate with 140 potential cluster stars lying within the inner 6 arcmin of the cluster, they identified background galaxies which were then used to determine the cluster's proper motion. The spread and uncertainties of those 16 background galaxies are large. Siegel et al.~find that this is due to a colour dependence of their proper motion, which is why they apply an ad-hoc colour correction to their sample and disregard 2 galaxies for which no colour information was available. From this corrected sample they derive the resulting proper motion and the relatively small uncertainties, which are also depicted in Fig.~\ref{propermotion}. 

In the same figure we also show the proper motion as artificially measured for all stars within the inner 6 arcmin of the three clusters with $M_0 = 5000\msun$ and $R_0 = 8$ pc. The stars from the cluster on \textit{orbit 1} lie exactly on the proper motion values of Siegel et al. (as expected since we used these values as input). Accordingly, the stars from the model on \textit{orbit 2} lie concentrated at zero proper motion. Only the stars of the cluster on \textit{orbit 3} show a significant intrinsic scatter of more than 0.5 mas pc$^{-1}$ about the central proper motion values of the cluster. This is due to extra-tidal stars which are on slightly different orbits than the cluster. Even though this scatter does not suffice to explain the discrepancy it may introduce an additional systematic uncertainty. Given that Siegel et al.~have not accounted for such a possible intrisic scatter, and given the nevertheless large scatter in their determined stellar proper motions (see their Fig.~5), we argue that in the curious case of Pal 13, the statistical and
systematic uncertainties in the proper motion determination may indeed
be (several times) larger than the formal error derived by Siegel et al.

A possible alternative would be that the Galactic potential which was used in our investigation does not properly reflect the true potential of the Milky Way. In order to keep the proper motion determined by \citet{Siegel01} and move Pal~13 closer to its apogalacticon we would have to modify the Galactic potential such that it is significantly stronger at the distance of Pal~13, as this would mean that the cluster gets decelerated more strongly and cannot get to large Galactocentric distances. But the most recent measurements of the Milky-Way potential agree well with our choice of potential, finding circular velocities of about 220 km/s at the distance of the Sun (e.g. \citealt{Ghez08}), and a possible flattening of the potential of about 0.87 \citep{Koposov10}. Such a flattening would imply the opposite of what would be necessary for bringing Pal~13 closer to its apogalacticon, since the Galactic force would be less strong at a given distance from the Galactic plane compared to the unflattened case.

\section{Summary and conclusions}\label{Sec:Conclusions}
We performed a set of $3\times15$ $N$-body computations of the low-mass Milky-Way globular cluster Palomar 13 which shows some peculiarities in observations. First of all, \citet{Cote02} measured a velocity dispersion of $2.2\pm0.4$ km/s, which yields a very high mass-to-light ratio of about 40 due to the cluster's low integrated absolute magnitude of only $M_V = -3.8$ mag. Secondly, Pal~13 shows a shallow slope at large radii within its surface density profile of $\eta = 1.9$, making a determination of its tidal radius difficult \citep{Siegel01, Cote02}. It has been suggested in both publications  that these effects might be due to Pal~13's last pericentre passage on its eccentric orbit about the Galactic centre. In contrast, \citet{Kuepper10b} find by means of $N$-body computations that pericentre passages barely influence the appearance of a cluster's surface density profile. Instead, they find a flattening of the surface density profile only for clusters on eccentric orbits which are about to reach their apogalacticon, resulting from the compression of the tidal debris as it gets decelerated in its orbit. 

We therefore use three different orbits to explicitly test these two hypotheses  for Pal 13 (Fig.~\ref{orbits}). First, we use the radial velocity measured by \citet{Cote02} in combination with the proper motion measured by \citet{Siegel01}. This yields an orbit in which Pal~13 today is close to its perigalacticon (\textit{orbit~1}). Secondly, we use only the measured radial velocity but set the proper motion to zero. This yields a similar orbit which is less eccentric and in which Pal~13 is today closer to apogalacticon but still not close enough for the effect of tidal debris compression described in \citet{Kuepper10b}, and which causes the surface density profile at large radii to become shallow, to take place (\textit{orbit~2}). Finally, we use the orbit with the proper motion which minimizes the orbital velocity of Pal~13, since this yields the orbit in which Pal~13 is today closest to apogalacticon (\textit{orbit~3}).

As it turns out, the model clusters on \textit{orbit~3} can readily reproduce Pal~13's peculiarities both in terms of surface density profile and velocity dispersion (and thus mass-to-light ratio), whereas the two other orbits cannot (Fig.~\ref{results}). While the first two orbits yield clusters with regular equilibrium velocity dispersion, the last orbit yields an enhanced velocity dispersions and a much larger spread in velocity dispersion values when measured from a subset of 21 stars. This is due to unbound stars within the cluster (potential escapers), and extra-tidal stars which get pushed back into the vicinity of the cluster when the cluster-tail system gets decelerated on its way to apogalacticon \citep{Kuepper10a, Kuepper10b}. 

With this investigation we would like to stress the importance of the orbital phase of a cluster on its appearance. Particularly interesting is the orbital phase just before reaching apogalacticon, where the cluster and its tails get decelerated and thus compressed such that the stellar density, especially in the region around the cluster's tidal radius, gets enhanced with unbound stars. These stars can alter the slope of the surface density profile at large radii, and increase the measured velocity dispersion significantly.

Since any cluster (or satellite in general) on an eccentric orbit about a galaxy spends most of its lifetime close to apogalacticon, it is likely to be observed in such a phase. It is therefore expected that a good fraction of all satellites are affected by this effect of orbital compression of their tidal debris. Observations not taking this effect into account may therefore assume too large tidal radii and/or ascribe a pronounced tidal debris to tidal shocking which in reality is only due to the deceleration of the satellite-tail system. And in some cases it may even lead to drastic overestimates of the dynamical mass, as is demonstrated here for Pal~13.

Moreover, orbital compression of a satellite's tidal debris can produce stellar systems, not only star clusters but also dwarf galaxies, which may appear largely extended, irregular in shape and dynamically hot, and thus may be misinterpreted, for instance, as bound systems embedded in dark matter haloes. Whether this effect can explain, for instance, the high mass-to-light ratios of dwarf galaxies such as Segue 1, which is currently known as the darkest ultra-faint dwarf galaxy \citep{Geha09}, has to be checked in a future investigation focussing on typical dwarf galaxy orbits and morphologies. Since dwarf galaxies are in general more diluted than globular clusters we expect that the effect is more pronounced in those cases. Available work in this direction by \citet{Kroupa97} and \citet{Klessen98} indeed suggests similar issues are relevant for dwarf spheroidal galaxies (see also \citealt{Kroupa10}). Segue 1, in fact, has recently been found in SDSS data to show a prominent tidal debris and therefore was re-classified as a dissolving star cluster \citep{Niederste09}. Its enhanced M/L ratio was interpreted as contaminated by stars from the Sagittarius stream but may well be due to the orbital compression effect described here.

Finally, this investigation poses the question how reliable proper motion measurement for halo satellites are (especially if they suffer from the orbital compression effect described in this work), or if we understand the potential of the Milky Way correctly, i.e. is the \citet{Allen91} potential which we used in this investigation a sufficient approximation? If the observed peculiarities in Pal~13 are indeed due to the orbital phase of the cluster, then either the proper motion measurement or the Galactic potential will be in question.

\section*{Acknowledgements}
The authors would like to thank Patrick C\^ot\'e for making his Palomar 13 data available. Moreover, they are grateful to Sverre Aarseth for making his \textsc{NBODY} codes accessible to the public. AHWK kindly acknowledges the support of an ESO Studentship and through the German Research Foundation (DFG) project KR 1635/28-1.

\bsp

\label{lastpage}
\end{document}